%% file: proceedingsblossierqcd12.tex
\documentclass[final,3p,times,twocolumn]{elsarticle}
 \biboptions{comma,sort&compress}
\usepackage{here}
\usepackage{axodraw}
 \usepackage{graphicx}
  \usepackage{epsfig}

\def\nin{\noindent}
\def\beq{\begin{equation}}
\def\eeq{\end{equation}}
\def\bea{\begin{eqnarray}}
\def\eea{\end{eqnarray}}
\def\nnb{\nonumber}
\def\la{\langle}
\def\ra{\rangle}

\newcommand{\ghvertex}{\begin{picture}(100,25)(0,-3)
\DashArrowLine(12.5,0)(50,0){5}
\DashArrowLine(50,0)(87.5,0){5}
\Gluon(50,0)(50,25){-4}{3}
\CCirc(50,0){5}{Black}{Yellow}
\Text(12.5,-10)[l]{k}
\Text(87.5,-10)[r]{q}
\Text(60,20)[l]{q-k}
\end{picture}}

\journal{Nuc. Phys. (Proc. Suppl.)}

\begin{document}

\begin{frontmatter}



\title{Lattice measurement of $\alpha_s$ with a realistic charm quark}

\author[label1]{B.~Blossier \corref{cor1}}
\cortext[cor1]{Speaker}
\ead{benoit.blossier@th.u-psud.fr}
\author[label1]{Ph.~Boucaud}
\author[label2]{M.~Brinet}
\author[label3]{F.~De Soto}
\author[label2]{X.~Du}
\author[label4]{V.~Mor\'enas}
\author[label1]{O.~P\`ene}
\author[label5]{K.~Petrov}
\author[label6]{J.~Rodr\'{\i}guez-Quintero}

\address[label1]{Laboratoire
de Physique Th\'eorique, Universit\'e
de Paris XI, B\^atiment 210, 91405 Orsay Cedex, France}

\address[label2]{Laboratoire de Physique Subatomique et de Cosmologie, CNRS/IN2P3/UJF; 
53, avenue des Martyrs, 38026 Grenoble, France}

\address[label3]{Dpto. Sistemas F\'isicos, Qu\'imicos y Naturales, 
Univ. Pablo de Olavide, 41013 Sevilla, Spain}

\address[label4]{Laboratoire de Physique Corpusculaire, Universit\'e Blaise Pascal, 
CNRS/IN2P3 63177 Aubi\`ere Cedex, France}

\address[label5]{Laboratoire de l'Acc\'el\'erateur Lin\'eaire -- IN2P3/CNRS,
Centre Scientifique d'Orsay, B\^atiment 200 -- BP 34, 91898 Orsay Cedex, France}

\address[label6]{Dpto. F\'isica Aplicada, Fac. Ciencias Experimentales; 
Universidad de Huelva, 21071 Huelva; Spain}


\begin{abstract}
\noindent
We report on an estimate of $\alpha_s$, renormalised in the
$\overline{MS}$ scheme at the $\tau$ and $Z^0$ mass
scales, by means of lattice QCD. Our major improvement compared to previous 
lattice calculations is that, for the first time, no perturbative treatment at 
the charm threshold has been required since we have used statistical samples of 
gluon fields built by incorporating the vacuum polarisation effects of 
$u/d$, $s$ and $c$ sea quarks. Extracting $\alpha_s$ in the Taylor
scheme from the lattice measurement of the ghost-ghost-gluon vertex, we 
obtain $\alpha^{\rm \overline{MS}}_s(m^2_Z)=0.1200(14)$ and 
$\alpha^{\rm \overline{MS}}_s(m^2_\tau)=0.339(13)$.

\end{abstract}




\end{frontmatter}


\section{Introduction}

\begin{figure}[hbt] 
\centerline{\includegraphics[width=6.cm,angle=90]{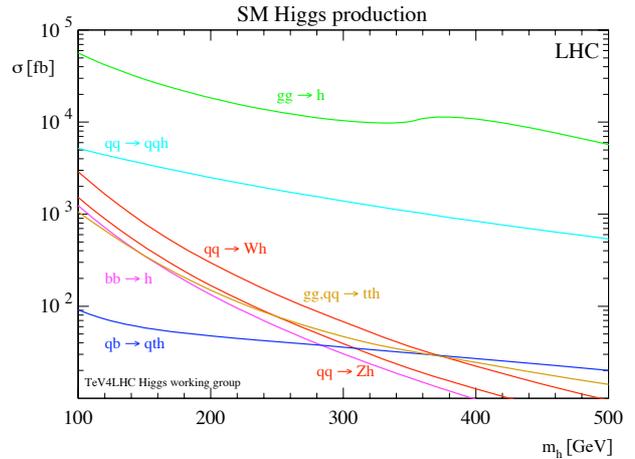}}
\caption{\scriptsize Prediction of the Standard Model BEH boson production in
function of its mass for different production channels at LHC}
\label{fig1} 
\end{figure}

\nin
The recent announcement by ATLAS and CMS of their observation at 5 $\sigma$
significance of a new particle with a mass around 125 GeV \cite{higgsATLAS}, interpreted 
as the Brout-Englert-Higgs (BEH) boson, makes even more crucial than before a satisfying 
control on theoretical inputs of analytical expression of the Higgs decay 
channels. Indeed, the era of precise Higgs physics (measurement of the 
couplings, ...) will certainly open soon: assessing the sensitivity of 
forthcoming detectors will be a key ingredient. There are different modes of
Higgs boson production: however the gluon-gluon fusion is by far the dominant
process, as shown in Fig.~\ref{fig1}. Over the uncertainty $\Delta 
\sigma^{\rm th}_{gg \to H \to X}$ of 20 - 25 \% claimed at LHC ($\sqrt{s} =
7\, {\rm TeV})$, about 4 \% come from the uncertainty $\delta \alpha_s$ on $\alpha_s
(m^2_{Z^0})$ \cite{Baglio}. A complementary approach of the $\alpha_s$ measurement
from the analysis of Deep Inelastic Scattering data, physics of jets, $\tau$
decay and $e^+ e^- \to$ hadrons \cite{reviewalpha} is its computation by 
numerical simulations. In the following section we will report on the work performed
by the ETM Collaboration to measure $\alpha_s$ from $N_f=2+1+1$ gauge configurations
\cite{ETMCalpha}.

\vspace{-0.4cm}
\section{$\alpha_s$ from numerical simulations}

\begin{figure}[hbt]
\centerline{\includegraphics[width=8.cm]{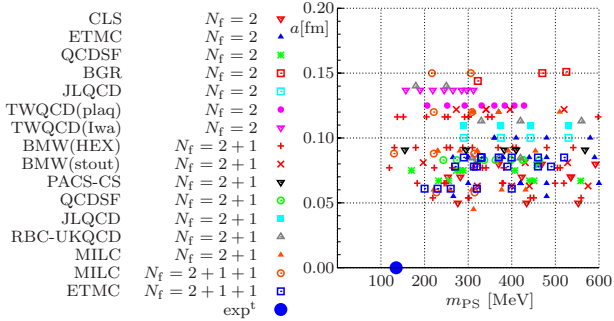}}
\caption{\scriptsize Simulation points obtained by the lattice community in the plane 
($a$, $m_\pi$)}
\label{figsimulations} 
\end{figure}

\nin
In the past years tremendous progresses have been made by the lattice community to
perform simulations that are closer to the physical point. It means including more and
more quark species in the sea, $u/d$ quarks $(N_f=2)$, then the strange $(N_f=2+1)$ and
even the charm $(N_f=2+1+1)$ since a couple of years. Pion masses $\sim 250$ MeV are now
common and several collaborations are even able to simulate a real pion, either in a
small volume (PACS-CS Collaboration) \cite{PACSCS} or using a quark regularisation with 
a rather aggressive cut-off of the UV regime (BMW Collaboration) \cite{BMW}. Discretisation 
errors are kept under control by considering lattice spacings $a$ smaller than 0.1 fm and 
lattice
extensions $L$ are such that $L m_\pi \gtrsim 3.5$ to get rid of finite size
effects. We have collected in Fig.~\ref{figsimulations} the simulation points performed by
the lattice community using different quark and gluon regularisations. 
\begin{figure}[hbt]
$F(p^2)$\\ 
\centerline{\includegraphics[width=6.cm]{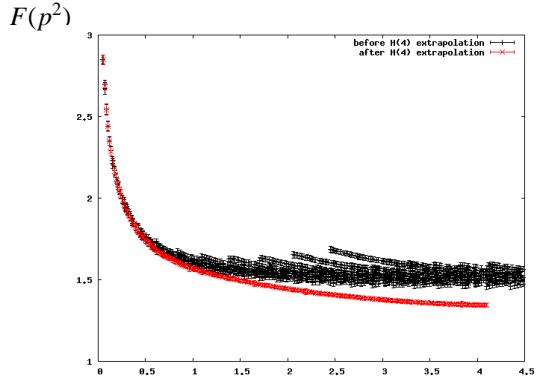}}
\caption{\scriptsize Raw and cured data of the ghost propagator dressing function}
\label{fig2} 
\end{figure}

\nin
The more vacuum polarisation effects are incorporated in the Monte-Carlo sample, 
the more reliable any result on $\alpha_s$ is. Several methods are proposed in the
literature to extract it: for instance analysing the static quark potential $V(r)$ at short
distance \cite{alpha1}, comparing the moments of charmonium 2-pt correlation function with
a perturbative formula after an extrapolation to the continuum limit \cite{alpha2},
integrating the $\beta$ function at discrete points in a finite volume renormalisation
scheme, for instance in the Schr\"odinger Functional scheme \cite{alpha3} or fitting
3-gluons amputated Green functions in the framework of Operator Product Expansion (OPE)
\cite{alpha4}. A last and particularly elegant approach consists in applying the OPE
formulae to the ghost-ghost-gluon amputated Green function \cite{alpha5}, that we will 
discuss in more details.\\
\nin
The starting point is to consider the bare gluon and ghost propagators in Landau gauge:
\bea\nnb
\left( G^{(2)} \right)_{\mu \nu}^{a b}(p^2,\Lambda)&=&
\frac{G(p^2,\Lambda)}{p^2} \ 
\delta_{a b} 
\left( \delta_{\mu \nu}-\frac{p_\mu p_\nu}{p^2} \right)\,,\\
\nnb
\left(F^{(2)} \right)^{ab}(p^2,\Lambda)&=&- \delta_{a b} \ 
\frac{F(p^2,\Lambda)}{p^2}\,.
\eea
Choosing a MOM scheme, the renormalised dressing functions $G_R$ and $F_R$, defined by
\bea\nnb
G_R(p^2,\mu^2)\ &=& \ \lim_{\Lambda \to \infty} Z_3^{-1}(\mu^2,\Lambda) \ 
G(p^2,\Lambda)\,,\\
\nnb
F_R(p^2,\mu^2)\ &=& \ \lim_{\Lambda \to \infty} 
\widetilde{Z}_3^{-1}(\mu^2,\Lambda)\ 
F(p^2,\Lambda)\,,
\eea
read $G_R(\mu^2,\mu^2)=F_R(\mu^2,\mu^2)=1$. The amputated ghost-gluon vertex is 
given by
\bea\nnb
\widetilde{\Gamma}^{abc}_\nu(-q,k;q-k) &=& 
\ghvertex \\
\nnb
\\
\nnb
&=&
i g_0 f^{abc} 
\left[ q_\nu H_1(q,k)\right.\\ 
\nnb
&&\left.+ (q-k)_\nu H_2(q,k) \right]\,.
\eea
The renormalised vertex is $\widetilde{\Gamma}_R=\widetilde{Z}_1 \Gamma$; with a 
MOM prescription it reads
\bea\nonumber
\lim_{\Lambda \to \infty}\widetilde{Z}_1(\mu^2,\Lambda)
\left.(H_1(q,k;\Lambda) +  H_2(q,k;\Lambda))\right\vert_{q^2=\mu^2}
=1.
\eea
The renormalised strong coupling constant is given by $g_R(\mu^2) =  \ 
\lim_{\Lambda \to \infty} g_0(\Lambda) \ \frac{Z_3^{1/2}(\mu^2,\Lambda)
\widetilde{Z}_3(\mu^2,\Lambda)}{ \widetilde{Z}_1(\mu^2,\Lambda)}$. In the case
of a zero incoming ghost momentum $k=0$, we are in a kinematical configuration where
the non renormalisation theorem by Taylor \cite{taylor} applies: 
$H_1(q,0;\Lambda)  +  H_2(q,0;\Lambda)=1$ and then
$\widetilde{Z}_1(\mu^2, \Lambda)=1$. The renormalised coupling in the
Taylor scheme reads finally
\bea\nnb
\alpha_T(\mu^2) \equiv \frac{g^2_T(\mu^2)}
{4 \pi}= \ \lim_{\Lambda \to \infty} \frac{g_0^2(\Lambda)}{4 \pi} 
G(\mu^2,\Lambda) F^{2}(\mu^2,\Lambda).
\eea
The main advantage of the MOM Taylor scheme is that there is no need to compute any 
3-pt correlation function: it is enough to extract the dressing functions of gluon and 
ghost propagators.

\nin
We have analysed the $N_f=2+1+1$ ensembles produced by the ETM Collaboration 
\cite{confETMC}, with 
bare couplings $\beta=2.1$, $1.95$ and $1.9$ that correspond to 
$a_{\beta=2.1} \sim 0.06$ fm, $a_{\beta=1.95}\sim 0.08$ fm and 
$a_{\beta=1.9}\sim 0.09$ fm, respectively. Pion masses are in the rang [250-325] MeV.
Landau gauge is obtained by standard methods to minimise $A^\mu A_\mu$
\cite{petrarcagiusti} while the ghost propagator is computed by inverting the
discretised Faddeev-Popov operator.
However, as the O(4) symmetry is broken on the lattice to the H(4) group, getting
$\alpha_T$ from the dressing functions $G$ and $F$ is not straightforward: ${\cal
O}(a^2 p^2)$ \emph{and} H(4) invariants artifacts, the so-called hypercubic artifacts,
have to be properly taken into account \cite{damirroiesnel}: 
\bea\nnb
\alpha_T^{\rm Latt}\left(a^2p^2,a^2 \frac{p^{[4]}}{p^2},\dots \right) 
\ &=& \ \widehat{\alpha}_T(a^2 p^2) \ \\
\nnb
&+& 
\left. \frac{\partial \alpha_T^{\rm Latt}}{\partial \left(a^2
\frac{p^{[4]}}{p^2}\right)}  
\right|_{a^2 \frac{p^{[4]}}{p^2}=0} \!\!\!\! 
a^2 \frac{p^{[4]}}{p^2} \ + \ \dots\,,
\eea
$p^{[4]}=\sum_i p_i^4$. We have shown in Fig.~\ref{fig2} that a "fishbone" structure, that are those hypercubic
artifacts, is clearly present in $F$ but curable, as also seen on the plot.
 \begin{figure}[hbt]
$\hat{\alpha}_T(p^2)$\\ 
\centerline{\includegraphics[width=6.cm]{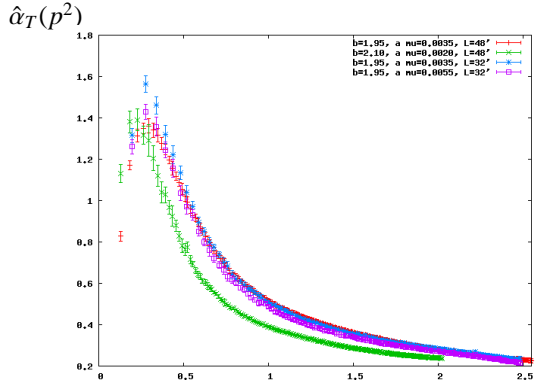}}
\caption{\scriptsize Strong coupling constant obtained after the elimination of the 
dominant hypercubic artifacts}
\label{fig3} 
\end{figure} 
\begin{figure}[hbt]
\centerline{\includegraphics[width=6.cm]{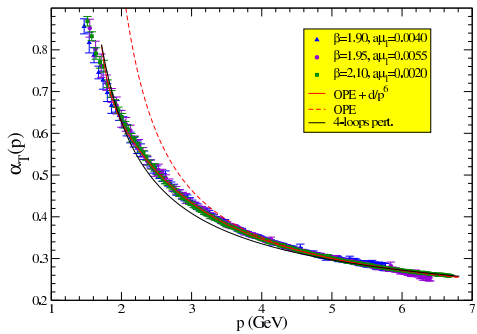}}
\caption{\scriptsize Raw data of $\alpha_T(p^2)$ compared to a purely perturbative
running and OPE with power corrections}
\label{fig4} 
\end{figure} 
\begin{figure}[hbt]
\centerline{\includegraphics[width=5.cm]{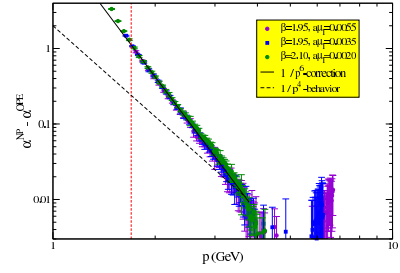}}
\caption{\scriptsize Subtraction of $\alpha_T(p^2)$ from the perturbative running
\newline
compared to generic power corrections.}
\label{fig5} 
\end{figure} 
\begin{figure}[!]
\centerline{\includegraphics[width=6.cm]{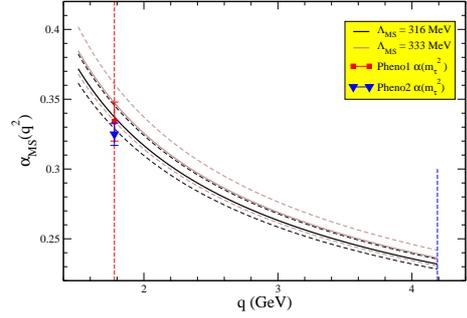}}
\caption{\scriptsize Comparison of $\alpha_s(m_\tau)$ measured on the lattice
with estimates 
\newline
based on phenomenological analyses of $\tau$ decay data 
\cite{taupich}, \cite{taunarison}.}
\label{fig6} 
\end{figure} 
The remaining cut-off effects are removed by fitting $\hat{\alpha}_T(p^2)$ 
according to the formula $\widehat{\alpha}_T(a^2 p^2) \ = \ \alpha_T(p^2) \ 
+ \ c_{a2p2} \ a^2 p^2 \ + \ {\cal O}(a^4)$. Fig.~\ref{fig3} illustrates the benefit, in
term of statistical error on $\alpha_T$, to do the calculation in Taylor scheme, as we 
pointed earlier in the text.\\
We then use the OPE formalism so that we can relate $\alpha_T(p^2)$ to 
$\alpha_T^{\rm pert}(p^2)$ \cite{alphaetmcNf2}, \emph{including power corrections}: 
$\alpha_T(p^2)= \alpha^{\rm pert}_T(p^2)
\left[  1 + \frac{C(p,q_0) 
g^2_T(q_0^2) \langle A^2 \rangle^{R}_{q_0^2}}{p^2} + 
\frac{d}{p^6} + \cdots \right]$, with $q_0$ fixed to 10 GeV and $C$ a combination of a Wilson
coefficient and a running of the gluonic operator $\la A^2 \ra$. Eventually 
$\alpha^{\rm pert}_T$ is
expressed at $\rm{N^3 LO}$ in function of $\Lambda_T$ with 
$\frac{\Lambda_{\overline{\rm MS}}}{\Lambda_T} \ \equiv 
 \exp\left(\displaystyle - \frac{507-40 N_f}{792 - 48 N_f}\right)$. The parameters to be
 fitted are thus $a \Lambda_{\overline{\rm MS}}$, $g^2_T \langle A^2 \rangle$, $d$ and the
 ratios of lattice spacings $a_{\beta}/a_{1.9}$ obtained by imposing that the various 
 curves of $\alpha_T$ merge onto a
 universal one.
We have shown in Fig.\ref{fig4} that the purely perturbative running formula does not
match with the $\alpha_T(p^2)$ data, adding the $1/p^2$ fits nicely with them down
to $p=3.5$ GeV while including the $1/p^6$ term improves our ability to describe them
further down to the $\tau$ mass scale. One could expect a $1/p^4$ power correction
but Fig.\ref{fig5} indicates that the fit is meaningless: still we do not exclude that
the corresponding Wilson coefficient would mimic an additional 
$1/p^2$ factor. Collecting in Tab.~\ref{tab1} $\Lambda_{\rm \overline{MS}}$, $g^2 \la A^2 \ra$ and 
the $d$ coefficient, with the lattice spacing $a_{1.9}=0.08612(42)$ fm \cite{confETMC}, we can 
run 
$\alpha^{\overline{\rm MS}}_s$ up to the $Z^0$ mass scale or at the $\tau$
mass scale. For the latter we obtain $\alpha^{\rm \overline{MS}}_s(m^2_\tau)=0.337(8)$
and  $\alpha^{\rm \overline{MS}}_s(m^2_\tau)=0.342(10)$ with our 2 estimates of 
$\Lambda_{\rm \overline{MS}}$; combining both of them and adding in quadrature the errors 
we get $\alpha^{\rm \overline{MS}}_s(m^2_\tau)=0.339(13)$. It is
in very good agreement with $\tau$ decay data 
analysed with dispersion relations \cite{taunarison}, \cite{taupich}, as plotted in 
Fig.\ref{fig6}. Running $\alpha_s$ up to the ${\rm \overline{MS}}$ scheme $b$ quark mass 
$m_b$, by using the $\beta$ function and the $\Lambda^{\rm N_f=4}_{\rm \overline{MS}}$
parameter we have measured, we can match with the ${\rm N_f}=5$ theory: $\alpha^{\rm N_f=5}_{\rm
\overline{MS}}(m^2_b) = \alpha^{\rm N_f=4}_{\rm \overline{MS}}(m^2_b) 
\left(1 + \sum_n c_{n0} (\alpha^{\rm N_f=4}_{\rm \overline{MS}}(m_b))^n \right)$. Then a second
running is applied up to the $Z^0$ mass scale. We obtain 
$\alpha^{\rm \overline{MS}}_s(m^2_{Z^0})=0.1198(9)$
and  $\alpha^{\rm \overline{MS}}_s(m^2_{Z^0})=0.1203(11)$ with, again, our 2 estimates of 
$\Lambda_{\rm \overline{MS}}$; combining both results and adding in quadrature the errors 
we get $\alpha^{\rm \overline{MS}}_s(m^2_{Z^0})=0.1200(14)$. We have shown in Fig.\ref{fig7} a 
comparison between 
lattice results \cite{alpha2}, \cite{alpha6} and \cite{ETMCalpha}, DIS data \cite{MSTW}, the world 
average quoted by the Particle Data Group \cite{alphaPDG} (WA '12) and a world average
realised by replacing the $N_f=2+1$ lattice results by the $N_f=2+1+1$ one (WA' '12), 
theoretically more reliable. In \cite{reviewalpha} one can find an almost exhaustive
collection of results. A single, very precise, lattice value dominates strongly the weighted
world average of $\alpha^{\rm \overline{MS}}_s(m^2_{Z^0})$; removing it enlarges its uncertainty.
Our estimates is in the same ballpark as other approaches and is using a complementary
framework.
{\scriptsize
\begin{table}[hbt]
\setlength{\tabcolsep}{0.4pc}
 \caption{\scriptsize   Fit parameters of $\alpha_T(p^2)$ analysed by means of OPE.}
    {\small
\begin{tabular}{|c|c|c|c|}
\hline
$\Lambda_{\overline{\rm MS}}^ {N_f=4}$ (MeV) & 
$g^2(q_0^2) \langle A^2 \rangle^{R}_{q^2_0} $ (GeV$^2$)&$d^{1/6}$ (GeV)&$\alpha(m_{Z})$\\ 
\hline
316(13) & 4.5(4)&&0.1198(9) \\
\hline
324(17) & 3.8(1.0)&1.72(3)&0.1203(11) \\
\hline
\end{tabular}
}
\label{tab1}
\end{table}
}

\vspace{-0.6cm}
\section{Conclusions}

\nin
We have reported on the first measurement of $\alpha_s$ from lattice simulations taking into
account the vacuum polarisation effects by charm quark in the so-called 
${\rm N_f}=2+1+1$ theory. The main benefit of our set-up is that there is no perturbative 
treatment at the charm threshold. We have used the OPE formalism to analyse gluon and ghost 
propagators to extract the strong coupling in the MOM Taylor scheme. We have taken care of the
hypercubic artifacts and included power corrections in the OPE, that cannot be neglected. 
An on-going project is to study whether other Green functions (3-gluon vertex, quark 
propagator,...) present the same feature: our extraction of $\Lambda_{\rm \overline{MS}}$
presented here will help us to reduce the uncertainty on the fits of those Green functions.\\
\\
K. Petrov acknowledges the support of "P2IO" Laboratory of Excellence.
\begin{figure}[t]
\vspace{1cm}
\begin{center}
\begin{picture}(47,-60)(47,-60)
\LinAxis(-60,-30)(240,-30)(5,4,-3.2,0,0.2)
\LinAxis(-60,-200)(240,-200)(5,4,3.2,0,0.2)
\Line(-60,-200)(-60,-30)
\Line(240,-200)(240,-30)

\Text(0,-210)[]{\small{0.116}}
\Text(60,-210)[]{\small{0.118}}
\Text(120,-210)[]{\small{0.12}}
\Text(180,-210)[]{\small{0.122}}

\Text(120,-230)[]{\small{$\alpha_s(m_Z)$}}

\Line(68,-48)(172,-48)
\Line(68,-46)(68,-50)
\Line(172,-46)(172,-50)
\CCirc(120,-48){3}{Red}{Red}

\Line(105,-68)(165,-68)
\Line(165,-66)(165,-70)
\Line(105,-66)(105,-70)
\CCirc(135,-68){3}{Blue}{Blue}

\Line(45,-88)(93,-88)
\Line(93,-86)(93,-90)
\Line(45,-86)(45,-90)
\CBoxc(69,-88)(4,4){Yellow}{Yellow}

\DashLine(-60,-95)(240,-95){2}

\Line(-19,-118)(88,-118)
\Line(-19,-120)(-19,-120)
\Line(88,-116)(88,-116)
\CBoxc(33,-118)(4,4){Blue}{Blue}

\DashLine(-60,-125)(240,-125){2}

\Line(51,-148)(93,-148)
\Line(93,-146)(93,-150)
\Line(51,-146)(51,-150)
\CBoxc(72,-148)(4,4){Green}{Green}

\Line(69,-168)(117,-168)
\Line(117,-170)(117,-166)
\Line(69,-166)(69,-170)
\CBoxc(93,-168)(4,4){Magenta}{Magenta}

\Text(-50,-40)[l]{${\rm N_f}=2+1+1$}
\Text(-50,-60)[l]{${\rm N_f}=2+1$}
\Text(-50,-80)[l]{${\rm N_f}=2+1$}
\Text(-50,-110)[l]{DIS}

\Text(184,-40)[l]{ETMC '12 }
\Text(184,-60)[l]{PACS-CS '11 }
\Text(184,-80)[l]{HPQCD '10 }
\Text(184,-110)[l]{MSTW '08}
\Text(184,-140)[l]{WA '12}
\Text(184,-160)[l]{WA' '12}

\end{picture}
\end{center}

\vspace{5.5cm}
\caption{\scriptsize Collection of results on $\alpha_s(m_Z)$.}
\label{fig7} 
\end{figure}
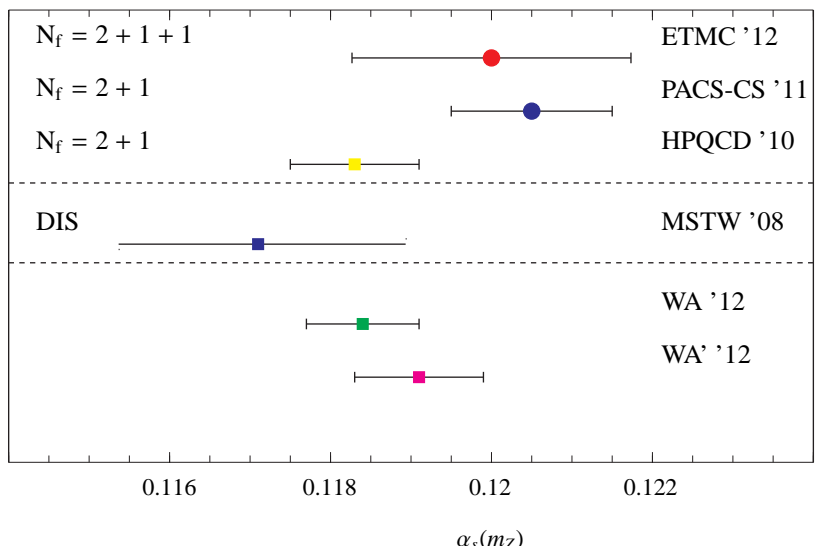 

\input{bib_proceedings}

\end{document}

%% file: bib_proceedings.tex

%% file: proceedingsblossierqcd12.bbl
\begin{thebibliography}{999}
\vspace*{-0.15cm}

\bibitem{higgsATLAS} ATLAS Collaboration, [arXiv:1207.7214]; CMS Collaboration, 
[arXiv:1207.7235].

\bibitem{Baglio} J. Baglio and A. Djouadi, JHEP {\bf 1103}, 055 (2011) 
[arXiv:1012.0530].

\bibitem{reviewalpha} S. Bethke \emph{et al}, [arXiv:1110.0016]. 

\bibitem{ETMCalpha} B. Blossier \emph{et al}, Phys. Rev. D {\bf 85}, 034503 (2012) 
[arXiv:1110.5829 [hep-lat]]; Phys. Rev. Lett. {\bf 108}, 262002 (2012) 
[arXiv:1201.5770].

\bibitem{PACSCS} S. Aoki \emph{et al}, Phys. Rev. D {\bf 81}, 074503 (2010) 
[arXiv:0911.2561].

\bibitem{BMW} S. D\"urr \emph{et al}, JHEP {\bf 1108}, 148 (2011) [arXiv:1011.2711].

\bibitem{alpha1} A. X. El-Khadra \emph{et al}, Phys. Rev. Lett. {\bf 69}, 729 (1992).

\bibitem{alpha2} I. Allison \emph{et al}, Phys. Rev. {\bf D 78}, 054513 (2008)
[arXiv:0805.2999].

\bibitem{alpha3} M. L\"uscher \emph{et al}, Nucl. Phys. {\bf B 389}, 247
(1993) [hep-lat/9207010]; Nucl. Phys. {\bf B 413}, 481 (1994) [hep-lat/9309005].

\bibitem{alpha4} B. Alles \emph{et al}, Nucl. Phys. {\bf B 502}, 325 (1997) 
[hep-lat/9605033]; Ph. Boucaud \emph{et al}, JHEP {\bf 9810}, 017 (1998) [hep-lat/9810322];
Phys. Rev. {\bf D 63}, 114003 (2001) [hep-ph/0101302].

\bibitem{alpha5} A. Sternbeck \emph{et al} PoS {\bf LAT2007}, 256 (2007) [arXiv:0710.2965]; 
Ph. Boucaud \emph{et al}, Phys. Rev. {\bf D 79}, 014508 (2009) [arXiv:0811.2059].

\bibitem{taylor} J. Taylor, Nucl. Phys. {\bf B 33}, 436 (1971).

\bibitem{confETMC} R. Baron \emph{et al}, JHEP {\bf 1006}, 111 (2010) [arXiv:1004.5284]; 
PoS {\bf LATTICE2010}, 123 (2010) [arXiv:1101.0518].

\bibitem{petrarcagiusti} L. Giusti \emph{et al},  Int. J. Mod. Phys. {\bf A 16}, 3487 (2001) 
[hep-lat/0104012]

\bibitem{damirroiesnel} D. Becirevic \emph{et al}, Phys. Rev. {\bf D 60}, 094509 (1999) 
[hep-ph/9903364]; F. De Soto and C. Roiesnel, JHEP {\bf 0709}, 007 (2007) [arXiv:0705.3523].


\bibitem{alphaetmcNf2} B. Blossier \emph{et al}, Phys. Rev. {\bf D 82}, 034510 (2010)
[arXiv:1005.5290].

\bibitem{taunarison} S. Narison, Phys. Lett. {\bf B 673}, 30 (2009) 
[arXiv:0901.3823].

\bibitem{taupich} A. Pich, [arXiv:1107.1123].

\bibitem{alpha6} S. Aoki \emph{et al}, JHEP {\bf 0910}, 053 (2009) [arXiv:0906.3906].

\bibitem{MSTW} A. Martin \emph{et al}, Eur. Phys. J. {\bf C 64}, 653 (2009) 
[arXiv:0905.3531].

\bibitem{alphaPDG} J. Beringer \emph{et al}, Phys. Rev. {\bf D 86}, 010001 (2012).


\end{thebibliography}
